\documentstyle[12pt]{article}
\input epsf.tex

\begin{document}

\baselineskip=15.5pt
\pagestyle{plain}
\setcounter{page}{1}

\begin{titlepage}

\begin{flushright}
PUPT-1814\\
hep-th/9810092
\end{flushright}
\vspace{15 mm}

\begin{center}
{\huge Baryons and String Creation\\
from the Fivebrane Worldvolume Action}
\end{center}

\vspace{12 mm}

\begin{center}
{\large Curtis G.\ Callan, Jr.\footnote{callan@viper.princeton.edu},
Alberto G\"uijosa \footnote{aguijosa@princeton.edu} and \\
Konstantin G. Savvidy \footnote{ksavvidi@princeton.edu} }\\
\vspace{3mm}
Joseph Henry Laboratories\\
Princeton University\\
Princeton, New Jersey 08544\\
\end{center}

\vspace{5 mm}

\begin{center}
{\large Abstract}
\end{center}
\noindent
We construct BPS-exact solutions of the worldvolume Born-Infeld 
plus WZW action of a D5-brane in the background of $N$ D3-branes. 
The non-trivial background metric and RR five-form field
strength play a crucial role in the solution. When a D5-brane is dragged 
across a stack of $N$ D3-branes a bundle of $N$ fundamental strings joining 
the two types of branes is created, as in  the Hanany-Witten effect. Our
solutions give a detailed description of this bundle in terms of a D5-brane 
wrapped on a sphere. We discuss extensions of these solutions which have an 
interpretation in terms of gauge theory multi-quark states
via the $AdS$/CFT correspondence.

\vspace{1cm}
\begin{flushleft}
October 1998
\end{flushleft}
\end{titlepage}
\newpage

% \renewcommand{\baselinestretch}{1.1}  %looks better
%%%%%%%%%%%%%%%%%%%%%%%%%%%%%%%%%%%%%%%%%%%
%%%
%% include the next line for double spacing %%
%%%%%%%%%%%%%%%%%%%%%%%%%%%%%%%%%%%%%%%%%%%
%%%
%\renewcommand{\baselinestretch}{2}

\newcommand{\grad}{\nabla}
\newcommand{\tr}{\mathop{\rm tr}}
\newcommand{\half}{{1\over 2}}
\newcommand{\third}{{1\over 3}}
\newcommand{\be}{\begin{equation}}
\newcommand{\ee}{\end{equation}}
\newcommand{\bea}{\begin{eqnarray}}
\newcommand{\eea}{\end{eqnarray}}
\newcommand{\ap}{\alpha^\prime}

\newcommand{\dint}[2]{\int\limits_{#1}^{#2}}
\newcommand{\D}{\displaystyle}
\newcommand{\PDT}[1]{\frac{\partial #1}{\partial t}}
\newcommand{\PD}{\partial}
\newcommand{\tw}{\tilde{w}}
\newcommand{\tg}{\tilde{g}}
\newcommand{\newcaption}[1]{\centerline{\parbox{6in}{\caption{#1}}}}
\def\href#1#2{#2}  

\section{Introduction}

In this paper we will construct solutions of the Born-Infeld action for a 
D5-brane in the background of a stack of $N$ D3-branes. 
By building on some recent work of Imamura 
\cite{Imamura}, we can find BPS-saturated solutions which presumably correspond
to exact solutions of string theory. Using the general approach of 
\cite{calmal,gibbons}, 
these solutions use D5-branes wrapped in various ways to describe branes and 
strings attached to each other. The primary object we construct this way is a 
D5-brane joined to $N$ D3-branes by a bundle of fundamental strings. Our 
solutions give a detailed description of the creation of these 
strings as the fivebrane is dragged across the threebranes.

As has been pointed out by various people \cite{wittenbaryon,groguri}, in the 
context of the anti-de~Sitter/conformal field theory ($AdS$/CFT)
correspondence there are general reasons to expect $N$ 
fundamental strings to join together on a D5-brane wrapped on a
five-sphere in the throat region of the threebrane geometry (i.e. in the $AdS$ 
geometry). This is the string theory counterpart to the gauge theory 
$SU(N)$ baryon vertex, representing a bound state of $N$ external quarks.
The baryon vertex has been studied in \cite{baryonsugra}, where the 
strings and the fivebrane are 
described in terms of separate Nambu-Goto actions. 
That approach ignores the worldvolume gauge field on the fivebrane. 
Its inclusion leads to the Born-Infeld action, which 
allows a unified description of the fivebrane and the strings. When 
restricted to the $AdS$ background, our solutions provide an explicit 
string theory representation of the baryon vertex. 
The Born-Infeld action for the worldbrane dynamics of the 
fivebrane in a threebrane background is an accessible and instructive way
to go after the energetics of this problem.

\section{The Setup}

We set up the equations for the Born-Infeld D5-brane in the background 
geometry of a stack of $N$ D3-branes. The metric in a standard 
coordinate system is
$$
ds^2=H(r)^{-1/2}(-dt^2+dx_{||}^2)+H(r)^{1/2}(dr^2+r^2 d\Omega_5^2), \qquad 
H(r)=a+R^4/r^4~.
$$
We have chosen to express $H(r)$ in terms of an auxiliary constant 
$a$, in order to treat
the asymptotically flat D3-brane ($a=1$)
and the $AdS_5\times {\rm \bf S}^{5}$ ($a=0$) geometries in parallel.
The worldvolume action is the Born-Infeld action calculated using the 
induced metric (including the worldvolume gauge field)
$$
g^{ind}_{\alpha\beta} = g_{MN}\partial_\alpha X^M \partial_\beta X^N + 
{\cal F}_{\alpha\beta},
$$
plus the WZW term induced by the five-form field strength; the
latter is basically a 
source term for the worldvolume gauge field. The explicit 
action we will use is
$$
S = -T_5\int d^6\xi\sqrt{-det(g^{ind})} +T_5 \int d^6\xi A_\alpha
        \partial_\beta X^{M_1}\wedge\ldots\partial_\gamma X^{M_5}G_{M_1\ldots 
M_5}~,
$$
where $T_5$ is the D5-brane tension and the second term is the explicit
WZW coupling of the worldvolume gauge field $A$ to the background 
five-form field strength $G$. We use the target space time and 
${\rm \bf S}^{5}$ spherical coordinates as worldvolume coordinates for the 
fivebrane, $\xi_{\alpha}=(t,\theta_{\alpha})$. 

We pick a five-sphere surrounding a point on the threebrane stack and look 
for static solutions of the form $r(\theta)$ and $A_0(\theta)$ (with all 
other fields set to zero), where $\theta$ is the polar angle in spherical 
coordinates. Non-static solutions are of interest too, but we will not
deal with them in this paper. 
On substituting explicit forms for the threebrane metric and the five-form 
field strength, the action (for static configurations) simplifies to
\be
\label{action}
S= T_5 \Omega_{4}\int dt d\theta \sin^4\theta \{ -r^4H(r)
  \sqrt{r^2+(r^\prime)^2 -F_{0\theta}^2}  +4 A_0 R^4 \},
\ee
where $\Omega_{4}=8\pi^{2}/3$ denotes the volume of the unit four-sphere.
 
The gauge field equation of motion following from this action reads 
$$
   (\sin\theta)^{-4} \partial_\theta \left[-\sin^4\theta 
{(a r^4+R^4)E\over\sqrt{r^2+{r^\prime}^2-E^2}}\right] =  4 R^4~,
$$
where we have set $E=F_{0\theta}$ and the right-hand side is the 
source term coming from the WZW
action. It is helpful to repackage this in terms of the displacement $D$ (the 
variation of the action with respect to $E$):
\be
\label{dispeq}
D={\sin^4\theta (a r^4+R^4)E\over \sqrt{r^2+{r^\prime}^2-E^2}}
\quad \Rightarrow \quad
     \partial_\theta D(\theta) = -4 R^4 \sin^4\theta.
\ee

Obviously, we can integrate the equation for $D$ to find it as an
explicit function of $\theta$. The result is
\be
\label{d}
D(\theta) = R^4\left[{3\over 2}(\nu\pi-\theta) 
  +{3\over 2}\sin\theta\cos\theta+\sin^{3}\theta\cos\theta\right],
\ee
where the integration constant has been written in terms of a parameter 
$0\le\nu\le1$, whose meaning will be elucidated below. 
Notice that the sign of 
the WZW term in (\ref{action}) reflects the choice of a particular 
fivebrane orientation. Choosing the opposite orientation 
therefore reverses the sign of the source term in (\ref{dispeq}), and 
consequently the sign of $D$.

Since $D$, unlike $E$, is completely unaffected by 
the form of the function $r(\theta)$, it makes sense to express the 
action in terms of $D$ and regard the result as a functional for
$r(\theta)$. It is best to do this by a Legendre transformation, 
rewriting the original Lagrangian as
$$
U =  T_5 \Omega_{4}\int d\theta\{D\cdot E+ \sin^4\theta (a r^4+R^4)
    \sqrt{r^2+(r^\prime)^2-E^2}\}~.  
$$
Integrating the $DE$ term by parts using $E=-\partial_\theta A_0$, 
one reproduces (with a sign switch) the original Lagrangian. 
Using (\ref{dispeq}) we can eliminate $E$ in favor of $D$ to get
the desired functional of $r(\theta)$ alone:
\be
\label{u}
U =  T_5 \Omega_{4}\int d\theta 
\sqrt{r^2+(r^\prime)^2}\sqrt{D^2+(a r^4+R^4)^2\sin^8\theta}.
\ee

This functional is reasonably simple, but complicated by
the fact that there is explicit dependence on $\theta$. Hence there
is no simple energy-conservation first integral that we can use to
solve the equations (or at least analyse possible solutions).
For future reference, we record the Euler-Lagrange equations that
follow from (\ref{u}):
\bea
\label{bigel}
{d\over d\theta}\Bigl({r^\prime\over\sqrt{r^2+{r^\prime}^2}}
     \sqrt{D^2+(a r^4+R^4)^2\sin^8{\theta}} \Bigr) =
 {r\over\sqrt{r^2+{r^\prime}^2}} \sqrt{D^2+(a r^4+R^4)^2\sin^8{\theta}}
      \nonumber \\
\hspace{4cm}
  +{\sqrt{r^2+{r^\prime}^2}\over r}{4 a r^4(a r^4+R^4)\sin^8\theta\over
\sqrt{D^2+(a r^4+R^4)^2\sin^8\theta}}~.
\eea
Supersymmetry considerations will allow us to go rather far in analysing the
solutions of this formidable-looking equation.

When we discuss solutions in more detail, we will see that
it will not be possible to wrap the fivebrane smoothly around a sphere. 
Even if $r(\theta)\sim r_0$ for most $\theta$, we will find that for 
$\theta\to \pi$ (or 0), $r$  shoots off to infinity in a way that simulates a 
bundle of fundamental strings in 
the manner described in \cite{calmal,gibbons}. 
Using (\ref{u})
we can already verify that the energy of such a spike is consistent with 
its interpretation as a bundle of strings. Suppose that the spike sticks out 
at $\theta=\pi$; then $D$ will take on some finite value $D(\pi)$ at
$\theta=\pi$. 
As we go into the spike, $r^\prime$ will dominate $r$ and the $D$ 
term will dominate $\sin^8\theta$. It is clear then that the spike has a 
`tension' (i.e. an energy per unit radial coordinate distance) 
$T_5 \Omega_{4} |D(\pi)|$.
Using the known facts that $D(\pi)=3\pi (\nu-1) R^4/2$ and 
$T_{5}\Omega_{4}R^{4}=2 N T_{f}/3\pi$, 
it follows that the `tension' of the spike is that of $n$ fundamental 
strings, $n T_f$, where $n=(1-\nu)N$. This gives meaning to the 
parameter $\nu$. 

\section{Supersymmetry Issues}

We are interested in placing a D5-brane in a D3-brane background and finding a 
structure
that looks like fundamental strings attached to the D3-branes. Insight into
what 
is possible 
is often obtained by looking for brane orientations such that the various
brane 
supersymmetry
conditions are mutually compatible for some number of supersymmetries. In type 
IIB
supergravity in ten-dimensional flat space, there are 32 supersymmetries 
generated by two
16-component constant Majorana-Weyl spinors $\eta_L,\eta_R$ of like parity
($\Gamma_{11}\eta_{L,R}=\eta_{L,R}$). In the presence of branes of various 
kinds, the 
number of supersymmetries is reduced by the imposition of further conditions. 
Explicitly,
\bea
\label{joe} \nonumber
{\rm F-string\ :} & \Gamma^{09}\eta_L=-\eta_L, & \Gamma^{09}\eta_R=+\eta_R, \\ 
{\rm D3-brane:} & \Gamma^{0123}\eta_L=+\eta_R, & \Gamma^{0123}\eta_R=-\eta_L,
\\ 
\nonumber
{\rm D5-brane:} & \Gamma^{045678}\eta_L=+\eta_R, & 
\Gamma^{045678}\eta_R=+\eta_L,
\eea
where the particular gamma matrix products are determined
by the embedding of the relevant branes into ten-dimensional space. For 
instance, the
D3-brane condition refers to a brane that spans the 123 coordinate directions. 
Conditions
can be multiplied by an overall sign by changing brane 
orientation.\footnote{Notice, however, that to maintain a 
supersymmetric configuration one must simultaneously reverse the orientation 
of two of the three objects.}
The existence of a BPS
state containing more than one type of brane depends on the existence of 
simultaneous
solutions of more than one of the above equations. The relevant point for our 
discussion is
that the conditions precisely as written above, corresponding to mutually 
perpendicular
D3-branes, D5-branes and F-strings, are compatible with eight supersymmetries 
(${\cal N}=2$ in 
usual parlance). The supersymmetry argument suggests that mutually orthogonal 
branes 
spanning a total of eight dimensions joined by a fundamental string running 
along the one
remaining dimension (perpendicular to both branes) should in fact be a stable 
BPS state.
An interesting aspect of our Born-Infeld worldvolume approach is that we will
explicitly see how the fundamental strings are created and destroyed as the 
D-branes are moved
past each other in the ninth direction (the Hanany-Witten effect 
\cite{hw, bdg, dfk, gilad}).

The above analysis has been carried out in flat space. To make contact with
the 
$AdS$/CFT correspondence, one would want to consider $N$ superposed D3-branes 
with $N$ large, in which case the background geometry is not flat and the 
supersymmetry analysis given above is at least incomplete. 
Imamura \cite{Imamura} has analysed the supersymmetry conditions 
associated with a D5-brane stretched over some surface in the `throat' of 
the D3-brane (where the geometry is
$AdS_5\times {\rm \bf S}^{5}$ and there is a flux of the RR five-form 
field strength
through the ${\rm \bf S}^{5}$). There are several new features here: first,
the 
unbroken supersymmetries of type IIB supergravity in this particular 
background are 32 position-dependent spinors (as opposed
to constant spinors in flat space); second, because of the RR five-form
field strength, there is a worldbrane gauge field induced on the 
worldvolume of the D5-brane; third, the condition that a particular element 
of the D5-brane worldvolume preserve some supersymmetry
involves the local orientation of the brane, the value of the induced 
worldvolume gauge field and the local value of the supergravity 
supersymmetry spinors. Since the D5-brane is embedded in
some nontrivial way in the  geometry, the supersymmetry condition is in 
principle different at each point on the worldvolume and it is far 
from obvious that it can be satisfied everywhere. 

However, Imamura \cite{Imamura} was able to show that these conditions boil
down, at least in the $AdS_5$ ($a=0$)
background, to a first-order equation for the embedding of the D5-brane into
the 
space
transverse to the D3-brane stack. In our language, his BPS condition can be 
written 
\be
\label{adsbps}
{r^\prime\over{r}} = {R^4\sin^5\theta +D(\theta)\cos\theta 
\over
     {R^4\sin^4\theta\cos\theta -D(\theta)\sin\theta} }~,
\ee
where $r=r(\theta)$ is the D5-brane embedding in the transverse space and 
$D(\theta)$ is the `displacement' field describing how the worldvolume 
gauge field varies from point to point. It is easy to show that any 
function $r(\theta)$ that satisfies this condition
automatically satisfies the Euler-Lagrange equation (\ref{bigel}) with
$a=0$; in that sense it is a first integral of the usual second-order equations.
Note that, as mentioned above, the structure of the action is such that there 
is no trivial energy first integral. The BPS condition (\ref{adsbps}) can be 
integrated analytically to obtain a two-parameter family of curves that 
describe BPS 
embeddings of a D5-brane into the $AdS_5\times {\bf S}^{5}$ geometry. 
These solutions 
will be discussed in the next section. 

We are also interested in exploring the analogous solutions in the
full asymptotically flat D3-brane 
background ($a=1$). In this background the 
interpretation and energetics of solutions should be quite straightforward. 
What is less obvious is how to find
BPS solutions. To follow Imamura's approach, one would first find the 
supersymmetry spinors in
the D3-brane background, use them to construct local supersymmetry conditions 
for an 
embedded D5-brane and from this find the condition on the embedding for there
to be a global worldvolume supersymmetry. 
This is no doubt perfectly feasible but we have not had
the patience to try it. Instead, we have simply guessed a generalization of
the $AdS_5\times{\bf S}^{5}$
BPS condition that automatically provides a solution of the Euler-Lagrange 
equations in
the full D3-brane background. The generalized BPS condition is obtained by 
making the (very plausible) replacement $R^4\to R^4+r^4$ in (\ref{adsbps}), 
\be
\label{d3bps}
{r^\prime\over{r}} = {(R^4+r^4)\sin^5\theta +D(\theta)\cos\theta 
\over
     {(R^4+r^4)\sin^4\theta\cos\theta -D(\theta)\sin\theta} }~.
\ee
It is easy to verify, using only (\ref{dispeq}), that this equation implies
the full Euler-Lagrange equation (\ref{bigel}) with $a=1$, so it is
certainly a first integral. Given its derivation, it is almost certainly
the BPS condition as well. It is rather surprising that things work 
so smoothly, and we take this as another evidence of the special nature 
of the D3-brane background. The first-order equation (\ref{d3bps})
must be integrated numerically (as far as we know) and yields a two-parameter 
family of solutions whose structure is quite non-trivial. Exploration of these 
and the $AdS$ solutions will be the subject of the rest of the paper.

Before closing this section we note that in either background
one can obtain an `alternative' BPS condition by reversing the signs in 
front of $D$ in equations (\ref{adsbps}) or (\ref{d3bps}). 
The resulting condition would guarantee the preservation of a 
different set of supersymmetries, albeit just as many. In order to 
still have a first integral of the Euler-Lagrange equation (\ref{bigel}), 
$D$ must satisfy (\ref{dispeq}) with the opposite sign for the source term. 
Such oppositely oriented fivebrane configurations will actually have
the same embedding $r(\theta)$. 

\section{Solution Overview and Interpretation}

\subsection{$AdS$ background: Born-Infeld Baryons}

We start with a discussion of the solutions of the $AdS$ BPS equation 
(\ref{adsbps}) for the supersymmetric embedding of a fivebrane 
in the $AdS_{5}\times {\rm \bf S}^{5}$ geometry, with topology 
${\rm \bf S}^{4}\times {\rm \bf R}$. Fortunately, the BPS equation 
has the following simple analytic solution:\footnote{We thank 
\O.~Tafjord for help in finding this solution.}
\be
\label{adssol}
r(\theta)=\frac{A}{\sin\theta}
          \left[\frac{\eta(\theta)}{\pi(1-\nu)}\right]^{1/3},
\qquad \eta(\theta)=\theta-\pi\nu-\sin\theta \cos\theta,
\ee
where the scale factor $A$ is arbitrary, and $\nu$ is the integration
constant in (\ref{d}). The freedom of changing 
$A$ is a direct consequence of the scale invariance 
of the $AdS$ background: if $r(\theta)$ is a solution of 
(\ref{adsbps}), then so is $\lambda r(\theta)$ for any $\lambda$. 
Note that $\eta>0$ (so that the solution makes sense) only for 
$\theta_{min}<\theta<\pi$, where $\theta_{min}$ is defined by 
\be
\label{nu}
\pi\nu=\theta_{min}-\sin\theta_{min}\cos\theta_{min}. 
\ee
This critical angle increases monotonically from zero to $\pi$
as $\nu$ increases from zero to one. Furthermore, when $\theta_{min}>0$,
$r(\theta_{min})=0$, a fact whose consequences will be explored 
below.
 
The fact that (if $\nu\neq 1$) the solution diverges as
$r\sim A/(\pi-\theta)$ when $\theta\to\pi$ 
means that a polar plot of $r(\theta)$ has, asymptotically, 
the shape of a `tube' of radius $A$. (This way of describing the
surface is a bit misleading as to the intrinsic geometry, but 
helps in visualization.) This tube is to be interpreted as a
bundle of fundamental strings running off to infinity in the space 
transverse to the D3-branes.
As explained in section 2, the asymptotic `tension' of the tube 
equals that of $(1-\nu)N$ fundamental strings. For the classical 
solutions $\nu$ is a continuous parameter, but at the quantum 
level $\nu$ should obey the quantization rule $\nu=n/N$.
  
In Fig.~1 we have plotted (\ref{adssol}) for some representative values
of $\nu$. Consider first the special case $\nu=0$, which yields a tube 
with the maximal asymptotic tension $N T_{f}$ and corresponds to the
classic `baryon' vertex. In this case the solution starts at a finite 
radius $r(0)=(2/3\pi)^{1/3}A$, with $r'(0)=0$, and then
$r(\theta)$ increases monotonically with $\theta$. 
The initial radius $r(0)$ represents another way of setting the overall 
scale of this scale-invariant solution. 
The fact that the fivebrane surface stays away from
the horizon at $r=0$ suggests that it is well-decoupled from degrees
of freedom living on the threebranes.

\begin{figure}[ht] \label{adstubes}                 
 \begin{center}
 \leavevmode
 \epsfbox{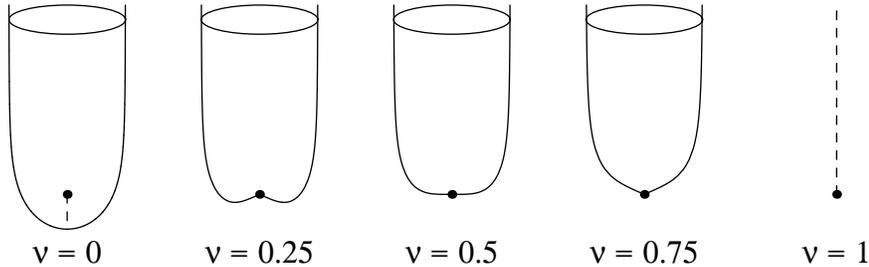}
 \end{center}
 {\small \caption[]{Polar plots of $r(\theta)$ for $AdS$ `tube' solutions 
   corresponding to $(1-\nu)N$ strings (with $\theta=\pi$ at the
   top of the plots). A 
  cross-section of each `tube' is an ${\rm \bf S}^{4}$.}}
\end{figure}

As far as the BPS equation is concerned, it seems to make sense to
consider the $\nu>0$ solutions as well. They have instructive features, 
although we will eventually conclude that they are on a less sound 
footing than their $\nu=0$ cousins. For large $r$, the solution asymptotes 
to the familiar tube with a tension corresponding to $(1-\nu)N<N$ strings:
it corresponds to a general multi-quark state of a $U(N)$ gauge theory. 
As mentioned above, for $\nu> 0$, the surface intersects $r=0$  
at an angle $\theta_{min}>0$ defined by (\ref{nu}), leading to the
cusp-like configurations displayed in Fig.~1. Note that, because 
the $r\to 0$ cusp has a finite opening angle, the fivebrane
does not capture all of the five-form flux: this is closely related
to the fact that the asymptotic tension is $(1-\nu)N$ and not $N$.

As $\nu\to 0$, the opening angle $\theta_{min}\to 0$. The approach 
to the $\nu=0$ solution, which does not contact $r=0$, is achieved, 
as shown in Fig.~1, via a `tensionless string' connecting the minimum
radius of the $\nu=0$ solution to $r=0$ (indicated as a dotted 
line in the figure). At the other extreme, $\nu\to 1$, one has 
$\theta_{min}\to\pi$, and the solution collapses to a similar 
`phantom string', this time running from $r=0$ to infinity.

One can compare the total energy of these configurations to that of 
$(1-\nu)N$ fundamental strings (this was done in \cite{Imamura} for 
the case of $\nu=0$). Using the solution (\ref{adssol}) in expression
(\ref{u}), the energy of the fivebrane up to an angular cutoff 
$\theta_{max}$ can be put in the form
\be
\label{uconfig}
U(\theta_{max})=NT_{f}\,\frac{A}{\pi}\int^{\theta_{max}}_{\theta_{min}}
   d\theta\,\left[\frac{\eta(\theta)}{\pi(1-\nu)}\right]^{1/3}
   \left\{\frac{\eta(\theta)}{\sin^{2}\theta}-\frac{4}{3}\sin\theta
   \cos\theta+\frac{4\sin^{4}\theta}{9\eta(\theta)}\right\}.
\ee
The fundamental string energy, on the other hand, for strings 
extending from the origin to a radial cutoff $r_{max}=r(\theta_{max})$,
is simply 
$E_{str}(\theta_{max})=(1-\nu)N T_{f}\,r(\theta_{max})$. 
It is easy to check numerically that  
$E_{str}(\theta_{max})-U(\theta_{max})\to 0$ as 
$\theta_{max}\to\pi$ ($r_{max}\to\infty$). 
The Born-Infeld fivebrane `tubes' can be therefore 
regarded as threshold bound states 
of $(1-\nu)N$ fundamental strings. We emphasize that this holds for any 
value of the scale parameter $A$: as $\theta_{max}\to\pi$, the energy 
$U(\theta_{max})$ becomes independent of $A$. The parameter $A$ is therefore 
a modulus in the space of equal-energy solutions. 

A complication for the interpretation of these solutions is that, in general
(specifically, when $\nu\ne 0,1/2,1$), the total five-form flux captured 
by the fivebrane differs from the number of fundamental strings, $(1-\nu)N$,
indicated by the asymptotic tension or total energy. The fundamental string 
charge is the source of the displacement field $D$, and we can rearrange 
(\ref{dispeq}) to show that a fivebrane that runs from $\theta=\theta_{min}$
to $\theta=\pi$ intercepts a total five-form flux
$$
Q_{flux}= -{2N\over 3\pi R^4}\left[ D(\pi)-D(\theta_{min})\right]
         = (1-\nu)N + {2N\over 3\pi} \sin^3\theta_{min} 
         \cos\theta_{min} ~.
$$
{}From the value of the tension, we would have expected a total 
charge $Q_{tot}=(1-\nu)N$ on the D5-brane. The difference,\footnote{
It should be clear that $Q_{flux}$, $Q_{tot}$, and 
$Q_{missing}$ all change sign if we reverse the fivebrane orientation.} 
\be
\label{qmiss}
Q_{missing}= -{2N\over 3\pi R^4} \sin^{3}\theta_{min}\cos\theta_{min}~,
\ee
is nonzero for $\nu\neq 0,1/2,1$ and presumably must be accounted for 
by a point charge at $r=0$. Since $r=0$ is where the fivebrane makes 
contact with the threebranes, this reminds us that, in order to be
fully consistent, we should take into account the possibility of exciting
the threebrane worldvolume $U(N)$ gauge fields when we attach fundamental
strings to the D3-branes (as in \cite{calmal,gibbons}). The case of $N$ 
strings ($\nu=0$) is special since they can be in a $U(N)$ singlet which 
will decouple from the D3-brane worldvolume gauge theory. When $\nu\ne 0$,
we are talking about a collection of strings that cannot be $U(N)$ neutral
and must excite the D3 gauge fields, which will in turn react back on the
metric. Since we have not allowed for this possiblity in our construction,
the detailed features of our solutions with $Q_{missing}\ne 0$ have to be 
taken with a certain grain of salt. The case $\nu=1/2$ is peculiar:
it corresponds to $N/2$ strings and so cannot form a $U(N)$ singlet, yet
has $Q_{missing}=0$. We are not sure that it really has the same
status as the true singlet $\nu=0$ solution. 

In light of the $AdS$/CFT correspondence \cite{jthroat,gkpads,wittenholo},
the above results are expected to have a gauge theory interpretation.
As discussed by several people \cite{wittenbaryon, groguri}, 
a baryon (a bound state of $N$ external quarks) in the $SU(N)$ ${\cal N}=4$ 
supersymmetric Yang-Mills (SYM) theory corresponds, 
in the dual $AdS$ description, to $N$ fundamental strings 
which join together on a D5-brane wrapped on an ${\rm \bf S}^{5}$ at some 
radius. The Born-Infeld $\nu=0$ fivebrane configuration described 
above provides a detailed representation of such a baryon. 
In particular, the absence of binding 
energy is as expected for a BPS threshold bound state in the
${\cal N}=4$ theory. Our other solutions with $\nu=n/N$ ($0<n<N$) 
are also BPS and correspond to threshold bound states of $N-n$ quarks. 
The existence of color non-neutral states with finite (renormalized) energy
is perfectly reasonable in a non-confining theory. To start learning
something interesting about these states, we would have to go beyond
mere energetics and ask some dynamical questions. 
To be absolutely clear, we emphasize that in every case discussed here
the quarks in the gauge theory are all at the same spatial location.

The solutions that we have discussed so far
are naturally restricted to the range of angles 
$\theta_{min}\leq\theta\leq\pi$ where $\eta(\theta)>0$. 
We will call them `upper tubes'. A simple modification
of (\ref{adssol}) is valid for the complementary angular range 
$0\leq\theta\leq\theta_{min}$ where $\eta(\theta)<0$:
\be
\label{adstildesol}
\tilde{r}(\theta)=\frac{\tilde{A}}{\sin\theta}
          \left[\frac{-\eta(\theta)}{\pi\nu}\right]^{1/3}.
\ee
This expression is singular at $\theta=0$
(where $\tilde{r}(\theta)\sim \tilde{A}/\theta$) and meets the origin at
$\theta=\theta_{min}$. It represents a downward-pointing tube of 
`radius' $\tilde{A}$ whose shape and tension are the same as those of 
an upward-pointing tube with parameter $1-\nu$. 
In other words, $\tilde{r}(\theta;\nu)=r(\pi-\theta;1-\nu)$.
This `lower tube' solution intercepts a total flux
$Q_{flux}=-\nu N + 2 N \sin^3\theta_{min}\cos\theta_{min}/3\pi$.
{}From the tension of this configuration, we would have expected a total
charge 
$Q_{tot}=-\nu N$, so there is a charge $Q_{missing}$ localized at the 
origin which is again given by (\ref{qmiss}). If the `upper tube' solution
corresponds to some number of quarks, the `lower tube' solution corresponds
to some number of antiquarks. 

Finally we want to speculate about constructing new solutions
by combining the $\nu\ne 0$ solutions we have been discussing.
Specifically, we are interested in obtaining configurations for which 
the peculiar charge at the origin cancels. 
Inspection of (\ref{qmiss}) shows 
that this can be achieved by merging two tubes whose opening angles 
are complementary. Using equation (\ref{nu}) this means that if one of 
the tubes has parameter $\nu$, the other one must have
parameter $1-\nu$. 
Taking into account the possibility of using `upper' or `lower' solutions, 
one is thus led to two types of configurations, illustrated
in Fig.~2. 

\begin{figure}[ht] \label{tubecomb}                 
 \begin{center}
 \leavevmode
 \epsfbox{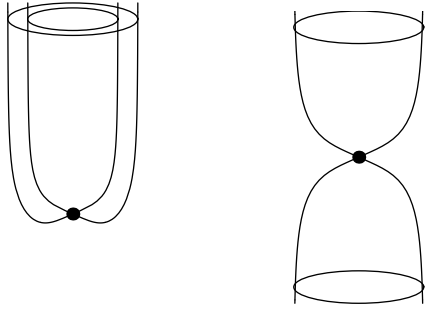}
 \end{center}
 {\small \caption[]{Upper/upper and upper/lower tube combinations. These 
  configurations have vanishing charge at the origin (see text).}}
\end{figure}

The combination of two upper tubes with parameters $\nu$ 
and $1-\nu$ yields a baryon-like configuration corresponding to 
a total of $N$ quarks. This system differs from the $\nu=0$ baryon 
of Fig.~1 in that the `strings' have been separated into two 
distinct coaxial tubes. 
It is interesting to note that this combined structure can be 
obtained as a single solution of (\ref{adsbps}), 
with a unique value of $\nu$, by formally continuing $r(\theta)$ 
in (\ref{adssol}) beyond 
$\theta=\theta_{min}$ (where $r=0$) to negative values of $r$.
The continued solution, depicted in Fig.~2,
can be interpreted as a single surface which
intersects itself at the origin.
In this interpretation the parameter $\nu$ is an additional modulus
of the baryon, controlling how many strings (out of the total of $N$)
`lie' in each tube.

If instead one puts together an upper $\nu$ and a lower $1-\nu$ 
solution, the result represents $1-\nu$ strings which 
extend from $r=\infty,\theta=0$ to $r=\infty,\theta=\pi$,
and run through the origin. 
In the gauge theory language this describes a state with 
$1-\nu$ quarks and the same number of antiquarks. 
(This is still BPS, because the quarks and antiquarks have 
opposite $SU(4)$ quantum numbers.) The total charge of the 
state vanishes.   

Judging from the cancellation of the charge at the origin, these 
combined solutions would appear to have the same status as the baryon. 
On the other hand, it is unclear to what extent these superposed 
tubes can be regarded as a single object, given that they are 
`connected' only at the horizon $r=0$. The issue is whether 
fluctuations can propagate from the lower tube to the upper
tube in finite `gauge theory' time. Since they would have to pass
through a horizon at $r=0$ to do so, this would appear to be
impossible, at least at the classical level. This issue merits
further study. 

\subsection{D3-brane background: Hanany-Witten effect}

So far, we have looked at the static solutions of D5-branes in the
$AdS_5\times {\rm \bf S}^5$ geometry of the `throat' region of 
the exterior geometry of a large number of D3-branes. As we 
now know, this limit gives us a supergravity description of 
${\cal N}=4$ SYM theory. We can also shed light on
some old string theory questions by studying the same types of 
configurations in the full asymptotically flat geometry of multiple
D3-branes.
 
To examine the character of the solutions in the asymptotically flat 
D3-brane background, it is convenient to parametrize the solution by
$z=z(\rho)$, where $\rho=r\sin\theta$, and $z=-r\cos\theta$. In these 
variables, adapted to flat space, the BPS condition (\ref{d3bps}) reads
\be
\label{d3bpsz}
z^\prime(\rho)=\frac{-D(\arctan(-\rho/z))}{\rho^{4}
               \left(1+\frac{R^{4}}{(\rho^{2}+z^{2})^{2}}\right)}\, .
\ee
Solutions to this equation describe D5-brane configurations which asymptote 
to a flat plane as $\rho\to\infty$ (equivalently, $\theta\to\pi/2$). 
The leading asymptotic behavior following from (\ref{d3bpsz}) is
\be
\label{asymptotplane}
z(\rho)=z_{max}+\frac{D(\pi/2)}{3\rho^{3}}
        +{\mathcal{O}}\left(\frac{R^{4}}{\rho^{4}}\right), 
\ee
where $z_{max}$ denotes the transverse position of the flat brane. 
We will be interested in studying how the solution changes 
as we vary $z_{max}$.
 
Figures 3 and 4
show the configurations obtained by integrating
(\ref{d3bpsz}) numerically for $\nu=0$ and $\nu=1/2$ and for a 
few representative values of $z_{max}$. The stack of $N$ D3-branes is at the 
origin, and extends along directions perpendicular to the figure.
For any value of $z_{max}$, the D5-brane captures the same fraction 
of the total five-form flux, which (in conjunction with a
possible point charge at the origin, as discussed in the previous
subsection) effectively endows the D5-brane with a total of 
$(1/2-\nu)N$ units of charge. Note the shift of $N/2$ units of charge, 
compared with the analogous situation in $AdS$ space: this happens 
because the asymptotic region of the brane is now at $\theta=\pi/2$
rather than $\theta=\pi$. This will have interesting consequences.

\begin{figure}[ht] \label{d3nuzero}                 
 \begin{center}
 \leavevmode
 \epsfbox{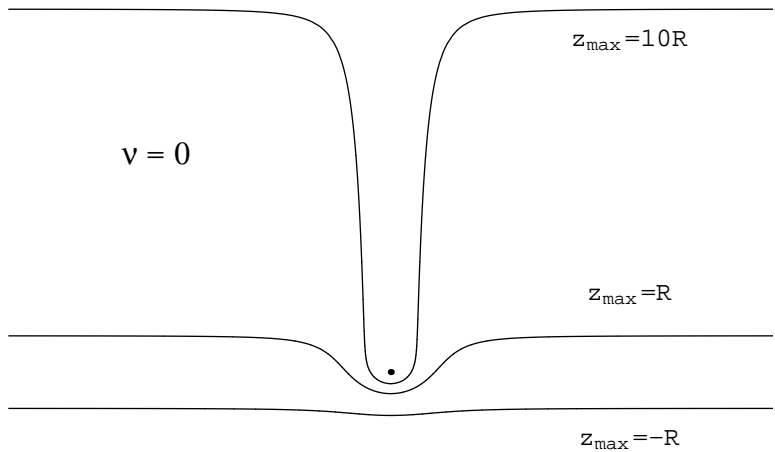}
 \end{center}
{\small \caption[]{Solutions describing the creation of $N$ fundamental 
   strings as a D5-brane is dragged upward, across a stack of 
   D3-branes. The number of strings connecting the two types of branes changes
   from 0 to $N$.}}
\end{figure}

Consider first the situation for $\nu=0$, described graphically in 
Fig.~3. 
As $z_{max}\to -\infty$, the charge density vanishes, and the D5-brane 
of course becomes flat. As one approaches the stack of threebranes 
from below ($z_{max}\to 0_{-}$), the charge becomes more and more 
localized near the center of the fivebrane, and the configuration becomes 
slightly deformed, bending away from the origin. 
As $z_{max}$ increases, the D5-brane remains `hung-up' on the 
D3-brane stack at $r=0$ and a tube of topology 
${\rm \bf S}^{4}\times {\rm \bf R}$ 
gets drawn out. The total charge of the tube itself approaches $N$
as it gets longer and longer and it becomes indistinguishable from a 
bundle of $N$ fundamental strings. Curiously, when the bundle eventually
connects to the flat D5-brane, a region of negative five-form flux is
encountered and the total charge intercepted by the fivebrane 
drops to $N/2$ (for any $z_{max}$). Altogether, then, this family 
of solutions provides a very concrete picture of the creation of 
fundamental strings as a fivebrane is dragged over a stack of 
threebranes, the Hanany-Witten effect \cite{hw, bdg, dfk,gilad}. 

\begin{figure}[t] \label{d3nuhalf}                 
 \begin{center}
 \leavevmode
 \epsfbox{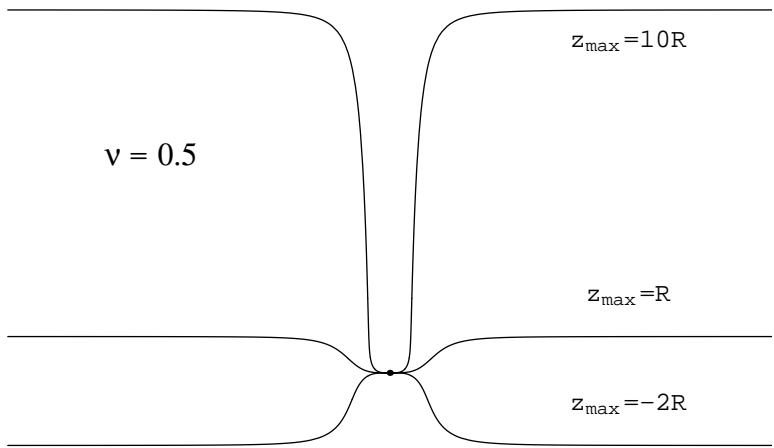}
 \end{center}
{\small \caption[]{Solutions describing the creation of $N$ fundamental strings
   as a D5-brane is dragged across a stack of 
   D3-branes. The number of strings connecting the two types of branes changes
   from $-N/2$ to $+N/2$ (the signs indicate whether the strings
   originate or terminate on the fivebrane).}}
\end{figure}

For $\nu>0$ the story is modified in exactly the same way as in 
the previous subsection. For either sign of $z_{max}$, the fivebrane 
now reaches the origin, $r=0$, at an angle $\theta=\theta_{min}$ 
given in terms of $\nu$ by equation (\ref{nu}). 
As $z_{max}\to -\infty$ the solution describes now a fivebrane 
connected by $\nu N$ strings to the stack of threebranes. 
For definiteness, assume the choice of sign for $D$ (i.e., the 
orientation of the fivebrane) is such
that the strings emanate from the D5-brane and terminate on the D3-branes.
Upon moving past $z_{max}=0$, $N$ strings 
directed towards the fivebrane are created, 
and as $z_{max}\to \infty$ $(1-\nu)N$ strings directed 
away from the threebranes extend between the two types of branes. The 
case $\nu=1/2$ is portrayed in Fig.~4.

It is instructive to compare the solution presented above to the 
description of fundamental strings attached to a fivebrane as
a Coulomb solution of the fivebrane Born-Infeld theory 
\cite{calmal,gibbons}. In the latter case the parent brane is 
embedded in flat space and the worldvolume gauge field is simply 
that of a point charge. For $n$ units of charge, the spike configuration
that protrudes from the brane at the location of the charge is of the form  
\be 
\label{cm}
z(\rho)=z_{max}-\frac{n c_{5}}{\rho^{3}},
\ee
where $c_{5}=2\pi^{2}g_{s}(\ap)^{2}$ is the quantum of charge. 
Writing this in terms of the threebrane throat radius
$R=(4\pi N g_{s}(\ap)^{2})^{1/4}$, one can readily verify that the 
asymptotic form (\ref{asymptotplane}) agrees with the solution
(\ref{cm}) for $n=(1/2-\nu)N$ strings. This is precisely as one would expect 
from the above discussion, for the entire fivebrane captures precisely 
a fraction $(1/2-\nu)$ of the total five-form flux. 
By the same token, it is clear 
that the present solution is of a more complex nature than that of 
\cite{calmal,gibbons}. The configuration discussed 
here corresponds roughly to a solution which is locally of the type 
(\ref{cm}), but where the charge $n$ varies as one changes position on 
the fivebrane.

One of the more confusing features of the Hanany-Witten effect is its
energetics: does the created string exert a force and, if so, how
is that consistent with the BPS property? We can shed some light on this by 
computing the energy of our configurations. In 
terms of the $z(\rho)$ parametrization, and using the fact that 
$T_{5}\Omega_{4}R^{4}=2 N T_{f}/3\pi$, equation (\ref{u}) becomes
\be
\label{uz}
U=N T_{f}\,\frac{2}{3\pi}\int d\rho\, 
\sqrt{1+\left(\partial_{\rho}z\right)^{2}}
\sqrt{D^{2}+\rho^{8}\left[1+\frac{R^{4}}{(\rho^{2}+z^{2})^{2}}\right]^{2}}\,.
\ee
We can use the BPS condition (\ref{d3bpsz}) to 
express the energy integrand
for our solutions solely as a function of $z$ and $\rho$,
\be
\label{uzrho}
U=N T_{f}\,\frac{2}{3\pi}\int d\rho\, 
 \left\{ \frac{D^2}{\rho^4\left[1+\frac{R^4}{(\rho^2+z^2)^2}\right]}
  +\rho^{4}\left[1+\frac{R^{4}}{(\rho^{2}+z^{2})^{2}}\right]\right\}.
\ee 

Now, the energy of the infinite D5-brane is evidently divergent, so 
(\ref{uzrho}) must be regularized. If we do so by placing a cutoff 
$\rho_{c}$ on $\rho$, the leading and subleading contributions
to (\ref{uzrho}) are clearly quintic and linear in $\rho_{c}$, 
respectively. 
The offending terms, however, are independent of 
$z_{max}$, so we choose to simply drop them (thereby removing an 
infinite constant from $U$). Altogether, then, we subtract 
$\rho^{4}+R^{4}$ from the integrand of (\ref{uzrho}) to obtain the
expression
\be
\label{usub}
\hat{U}(z_{max},\rho_{c})=N T_{f}\,\frac{2}{3\pi}\int^{\rho_{c}}_{0}d\rho\, 
 \left\{\frac{D^2}{\rho^4\left[1+\frac{R^4}{(\rho^2+z^2)^2}\right]}
  -\frac{2\rho^{2}z^{2}+z^{4}}{(\rho^{2}+z^{2})^{2}}\right\},
\ee 
which is finite as the cutoff is removed. Using the numerical solutions, 
one finds that in fact
$\hat{U}\to (1/2-\nu) N T_{f}\, z_{max}$ 
for all $z_{max}$ as $\rho_{c}\to\infty$.

Since the difference between $U$ and $\hat{U}$ is a constant, it 
readily follows that there is a net constant force on the fivebrane,
independent of $z_{max}$:
\be
\label{force}
\frac{\partial}{\partial z_{max}}U(z_{max},\rho_{c}) \to
       \left(\frac{1}{2}-\nu\right) N T_{f}
\quad\mbox{as}\quad \rho_{c}\to\infty.
\ee
The total force equals the tension of $(1/2-\nu)N$ 
fundamental strings as a consequence of the fact that the full
configuration carries a total of $(1/2-\nu)N$ units of charge. It might seem 
surprising at first that there is a force on the fivebrane 
even for $\nu=0$ and $z_{max}<0$ 
(i.e., before the D5-brane crosses the 
D3-brane stack and a tube of strings is created),
but one must keep in mind that even then there is 
a charge on the brane, and consequently a position-dependent energy.
After the fivebrane is moved past the threebranes, to $z_{max}>0$, a 
bundle of fundamental strings is created, and this bundle pulls down 
on the fivebrane with a force which tends to $N T_{f}$ as 
$z_{max}\to\infty$. The net force on the brane is still $(1/2-\nu)N T_{f}$, 
however, because the outer portion of the brane now carries negative 
charge, as a consequence of which there is an additional, upward force 
on the brane.
Our approach thus makes it absolutely clear that, contrary to the naive
expectation, there is no discontinuity in the force as the branes cross.
One is able to find static solutions despite the 
presence of a constant force because the D5-brane is infinitely 
massive, and therefore will not move.
If desired, this constant force can be cancelled by placing 
$(1-2\nu)N$ additional D3-branes at $z=-\infty$.

An alternative way to reach the same conclusions is to compute the 
force by cutting off (\ref{uz}) at $\rho_{c}$ and differentiating with 
respect to $z_{max}$ under the integral (regarding 
$z=z(\rho;z_{max})$). After an integration by 
parts and an application of the Euler-Lagrange equation, one is left 
only with a boundary term, which yields the analytic expression
\be
\label{slope}
\frac{\partial U}{\partial z_{max}}(z_{max},\rho_{c})=
N T_{f}\,\frac{2}{3\pi}\left.\left\{ 
\frac{\partial_{\rho}z}{\sqrt{1+(\partial_{\rho}z)^{2}}}
\sqrt{D^{2}+\rho^{8}\left[1+\frac{R^{4}}{(\rho^{2}+z^{2})^{2}}\right]^{2}}
\frac{\partial z}{\partial z_{max}}\right\}\right|_{\rho_{c}}.
\ee
Using (\ref{slope}) and (\ref{asymptotplane}) one can again conclude 
that the force on the fivebrane approaches $(1/2-\nu)N T_{f}$ as 
$\rho_{c}\to\infty$ at a fixed $z_{max}$. Furthermore, 
using the numerical solution in conjunction with (\ref{slope}), one 
can compute the force on the $\rho<\rho_{c}$ portion of the brane, 
for any value of $\rho_{c}$. For any fixed $\rho_{c}$, it is easy to 
see that the force tends to $(1-\nu)N T_{f}$ ($\nu N T_{f}$)
as $z_{max}\to\infty$ ($z_{max}\to-\infty$). Taking the limit this
way picks out the stress on the string tube part of the configuration
and yields the expected tension of $(1-\nu)N$ ($\nu N$) fundamental 
strings.
Nonetheless, the total asymptotic stress on the D5-brane is smaller 
by $N/2$ fundamental string units, due to the extra five-form 
flux intercepted by the flat part of the D5-brane. 

Notice that for $\nu=1/2$ the net force on the D5-brane vanishes. This is 
because in that case the total charge on the 
brane is zero. As a result, the $\rho^{-3}$ term in 
(\ref{asymptotplane}) has a vanishing coefficient.
This configuration has $z( \rho=0) = z'(\rho=0)=0$,
$\theta_{min}= \pi/2$, and $D(\theta_{min}) =0$. It is only for this
and the $\nu=0$ and $\nu=1$ cases that
the point charge at the origin vanishes.
This solution (depicted in Fig.~4)
describes a bundle of $N/2$ strings which flip their orientation
as the fivebrane to which they are attached is moved 
above or below the threebrane stack.
The number of attached strings still changes by $N$,
from $-N/2$ to $+N/2$, as the D5-brane is pulled through the
stack.  This configuration is thus a realization 
 of the `half-string' ground-state of the system described in 
\cite{dfk, gilad}. For reasons explained in those papers, and confirmed
by our energy analysis, this is the only solution which is in a
state of neutral equilibrium.

\begin{figure}[ht] \label{d5d5strings}                 
 \begin{center}
 \leavevmode
 \epsfbox{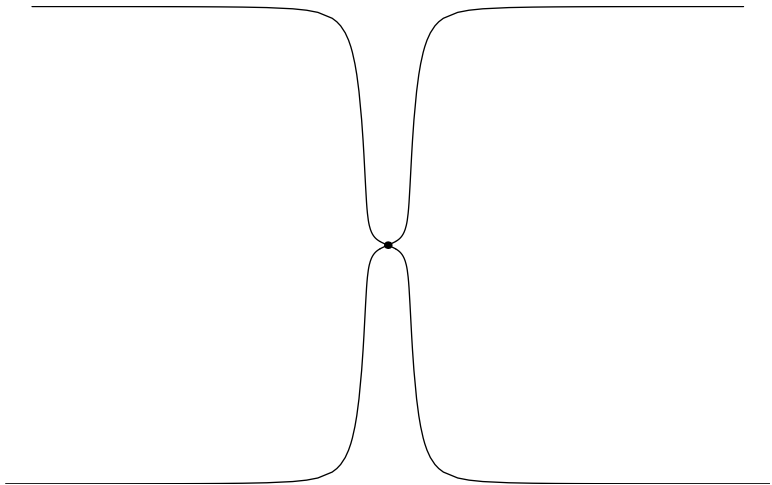}
 \end{center}
{\small \caption[]{Solution describing a system of two parallel D5-branes
   connected by $(1-\nu)N$ fundamental strings which run through the 
   D3-branes at $r=0$. A `point W-boson charge' lies at the origin.}}
\end{figure}

Just as in the previous subsection, one could imagine combining solutions
to obtain configurations in which the charge at the origin vanishes.
We will focus attention here on the possibility of superposing 
two solutions with parameters 
$\nu$, $z_{max}>0$ and $1-\nu$,$-z_{max}$, respectively. 
The complete structure obtained
this way is illustrated in Fig.~5, and corresponds
to a configuration in which two infinite parallel D5-branes with the same 
orientation, located at $\pm z_{max}$, are connected by $(1-\nu)N$ 
fundamental strings running through the $N$ D3-branes at the origin. 
Something interesting has happened here: we have constructed an
excitation of a system of two parallel fivebranes, something which
should more properly be described by the non-abelian $SU(2)$
Born-Infeld action. We have achieved this effect (perhaps illegitimately!)
by gluing together two $U(1)$ solutions at the singularity provided
by the D3-branes. The result is reminiscent of the configurations
examined in \cite{aki}, where the SYM Prasad-Sommerfield monopole 
solution is used to describe a system of 
two parallel D3-branes connected by a string. 
Following that 
interpretation, the point charges of the component
solutions at the origin should be understood not to cancel, but 
to combine instead into a `point W-boson charge' which interpolates,
at no cost in energy, between the two $U(1)$s of the overall broken
$U(2)$ symmetry. 

The threebrane background geometry evidently plays a role
in facilitating the construction described above. Nevertheless,
since the `strings' in the solution are interpreted as 
merely passing through  
the D3-branes, it is natural to conjecture that there exist
neighboring static fivebrane configurations in 
which the strings miss the origin. Deforming the system in this manner 
it would be possible to move the connecting strings arbitrarily far 
away from the threebranes, thereby producing the analogous flat 
space configuration. It would be very interesting to pursue this issue 
further.

\section{Conclusions}

It is surprising how many subtle aspects of the dynamics of branes
and strings can be illuminated by the Born-Infeld worldvolume gauge
theory approach. The principal focus of this paper has been the 
phenomenon of string creation when fivebranes are dragged across 
threebranes, but there are other issues which we did not discuss here,
and which might repay study. It would be straightforward to look at the
structurally very similar non-BPS configurations of D5-branes in 
non-extremal D3-brane backgrounds. Via the $AdS$/CFT correspondence, this
would tell us about multi-quark states in confining gauge theories. 
It would also be very instructive to study the case of multi-center 
BPS D3-brane
configurations in order to understand the effects of Higgs breaking
of the underlying $U(N)$ gauge symmetry. We hope to pursue these and
other subjects in future work. 

\section*{Acknowledgements}

This work was supported in part by US Department of Energy 
grant DE-FG02-91ER40671 and by National Science Foundation grant
PHY96-00258. AG is also supported by the National Science and 
Technology Council of Mexico (CONACYT).
CGC acknowledges the hospitality of the Aspen Center for
Physics where some of this work was done. He would also like to thank
J.~Maldacena for several insightful suggestions. AG is grateful to 
I.~Klebanov and {\O}.~Tafjord for many useful discussions. KGS would like 
to thank G.~Savvidy for numerous helpful conversations.

\end{document}